# Controlling Magnetization of Gr/Ni Composite for Application in High Performance Magnetic Sensors

M. R. Hajiali[1,*], L. Jamilpanah[1], Z. Sheykhifard[1], M. Mokhtarzadeh[1], H. F. Yazdi, B. Tork[1], J. Shoa e Gharehbagh[1], B. Azizi[1], S. E. Roozmeh[2], G. R. Jafari[1] and Majid Mohseni[1,*]

[1] Faculty of Physics, Shahid Beheshti University, Evin, 19839 Tehran, Iran
[2] Department of Physics, University of Kashan, 87317 Kashan, Iran

**ABSTRACT**

Graphene (Gr), a well-known 2D material, has been under intensive investigation in the last decade due to its high potential applications in industry and advanced technological elements. The Gr, while composed with magnetic materials, has opened new opportunities for further developments of magnetic based devices. Here, we report a mass production of Gr/Ni composite powders using electrochemical exfoliation/deposition method with different magnetic strengths of the final composite material. We applied the magnetic composite materials in a magnetoimpedance (MI) based sensor and observed significant enhancement in the MI effect and its field sensitivity. Such magnetic composites with controlled magnetization strengths are coated on the MI-ribbon sensor surface and different MI responses are observed. The MI response of a ribbon coated with a Gr/Ni layer is theoretically determined based on an electrodynamic model with a qualitative consistency between the experimental results and the theoretical model. Our comprehensive study can be applied in high performance functionalized MI based magnetic sensors and devices.

## 1. INTRODUCTION

In recent years, graphene, a two-dimensional (2D) hexagonal lattice of sp² hybridized carbon atoms, has attracted comprehensive research interest because of its fascinating electrical, mechanical, chemical and optical properties and its potential application in next-generation electronic [1,2], energy storage devices [3,4,5] and composite based materials [6–8]. Integration of graphene or graphene family with various nanoparticles allows the development of new nanocomposite materials with novel properties and highly promising applications in bioscience [9–11], microwave elements [12,13], sensors, [14] etc.[15].

The combination of graphene with nanoparticles, thereby forming graphene−nanoparticle hybrid structures, offers a number of additional unique physiochemical properties and functions that are both highly desirable and markedly advantageous for various applications when compared to the use of either material alone [16]. These graphene− nanoparticle hybrid structures are especially alluring because not only do they display the individual properties of the nanoparticles which can already possess beneficial optical, electronic, magnetic, and structural properties that are unavailable in bulk materials, but they also exhibit additional advantages and often synergistic properties that greatly augment their potential use for bio-applications [17]. On the other hand, in sensing applications, the combination of magnetic nanoparticles (MNPs), with alternative functionalization and catalytic properties, with graphene materials allows for the enhancement of sensitivity and selectivity over either graphene or nanoparticle magnetic based sensors alone [16–19].

A wide range of magnetic sensors, such as anisotropic magnetoresistance, giant magnetoresistance, tunneling magnetoresistance, the Hall effect and magnetoimpedance (MI) sensors are now available. Among them, MI effect has been considered as an effect with higher field sensitivity and appropriate signal intensity for magnetic sensors [20–22]. The MI effect is defined as the change of the electrical impedance of a conducting FM with high transverse magnetic permeability ($\mu_t$) in the presence of a static magnetic field [23–29]. By applying an external magnetic field, the skin depth ($\delta$) changes due to the change in $\mu_t$, thus varying the impedance of the FM. In the case of the ribbon, with width $l$ and length $L$, the impedance is approximately

$$Z = (1-i)\frac{\rho L}{2l\delta} = \frac{(1-i)L}{2l}(\pi \rho f \mu_t)^{\frac{1}{2}}$$

(1)

where $\rho$ is electric resistivity, $f$ is frequency of the current and $i$ = imaginary unit. Therefore, the impedance of the ribbon is a function of frequency, driving current and the external dc magnetic field ($H_{dc}$) through $\mu_t$ and $\delta$. This phenomenon has two aspects, one is related to improving the magnetic field sensor performance, and another one is related to the environmental sensing ability. There are reports on the magnetic field sensitivity enhancement by coating layers with different magnetization and conductivities on the surface of MI sensors [28,30–34]. The physical origins are well known through magnetic proximity effect [8], decrease of surface roughness and magnetic exchange interactions between sensors and coated magnetic layers. Also, surface modification of the ribbons has been reported intensively for detection of environmental elements through magnetic and nonmagnetic interactions with the environment [21].

Recently, we applied electrochemical exfoliation of graphene using a solution containing Ni ions which leads to simultaneous exfoliation of Gr and deposition of Ni on Gr sheets [15]. Here, we extend this method for fabrication of samples with controlled amount of magnetization of the samples and introduce a useful technological application of such magnetic Gr/Ni composites. We investigate the application of the synthesized materials in MI sensors with observed enhanced ratio and sensitivity when applied atop an MI sensor.

In order to obtain a comprehensive study and description of the experimental results, we proposed an electrodynamic model to describe the MI in a ribbon covered by solving the linearized Maxwell equations for the electromagnetic fields coupled with the Landau-Lifshitz-Gilbert (LLG) equation for the magnetization dynamics of the sensor. These results can be used in interpreting the experimental results of MI based sensors. Both theoretical predictions and experimental observations confirm the MI response quality change due to coating with magnetic composites with different magnetization strengths. The fundamental physical phenomenon described in this work can be used for development of magnetic field and environmental sensors based on the MI effect.

## 2. SAMPLE PREPARATION AND EXPERIMENTAL METHODS

### 2.1. Preparation of Gr/Ni Composite by Electrochemical Exfoliation

Electrochemical exfoliation of graphite was performed in a two-electrode system using platinized silicon (100 nm thickness and lateral dimension of $0.5 \times 10$ cm$^2$) as the cathode electrode and a graphite foil ($2 \times 10$ cm$^2$) as the anode electrode. The distance between the two electrodes was kept constant at 2.7 cm. Electrolyte solutions were prepared by dissolving $NiCl_2.6H_2O$ powder in water with three different molar ratios of 0.05, 0.075 and 0.1. A constant voltage (+10 V) was applied to the electrodes to provide expansion, exfoliation of graphite and deposition of Ni. The voltage was maintained constant for 20 min to complete the exfoliation/deposition process. Afterward, the product was collected using vacuum filtration and repeatedly washed with DI water. The resulted product was dispersed in water for sonication.

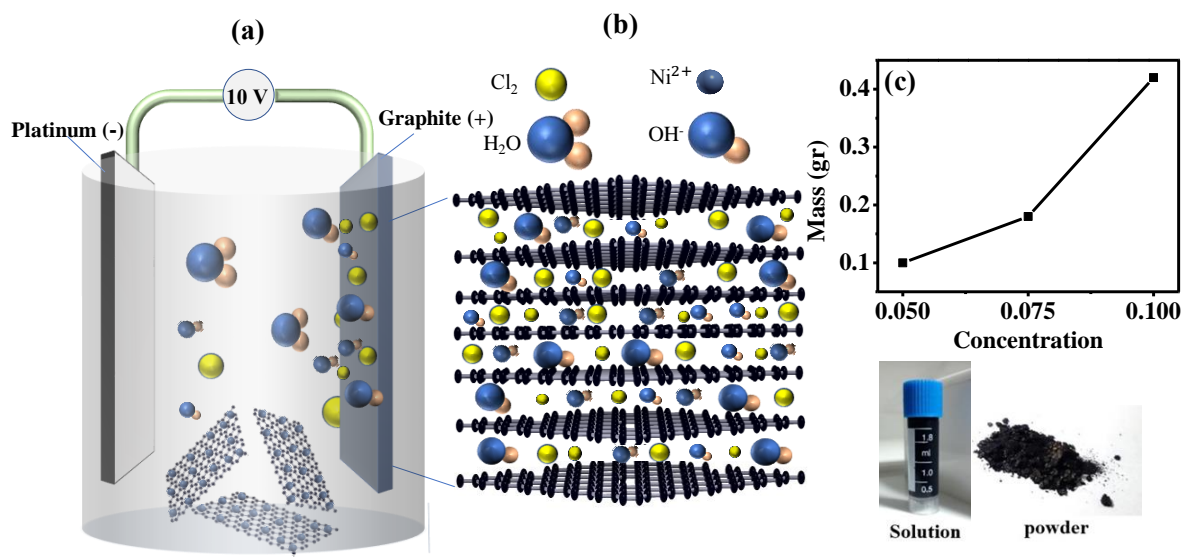

**Figure 1.** (a) Schematic illustration of the proposed mechanism for magnetic Gr/Ni production, (b) Situation of OH⁻ and Cl⁻ ions between graphite sheets which increases the interlayer distance. Cl gas can exert excessive force to graphite layers which results in separation of the graphite layers and then Gr sheets distributed in the solution can trap $Ni^{2+}$ ions. (c) Concentration of electrolyte solution and final mass of magnetic Gr/Ni composite.

The mechanism of electrochemical exfoliation is depicted in Figure 1a, b. First, by applying a voltage between the electrodes, hydroxyl ions (OH⁻) are produced in the cathode region and then accelerate towards the anode and hit the graphite surface. The collision of graphite with OH⁻ ions initially occurs at each side and grain boundaries. The oxidation at the edge side and grain boundaries leads to the expansion of the graphite layers; therefore, Cl⁻ ions can penetrate through them and the reduction of Cl⁻ ions produce Cl gas. The gas can exert an excessive force to graphite layers which results in separation of the layers [35]. In continue, the distributed Gr sheets in the solution can trap $Ni^{2+}$ ions. Since such sheets have been partially charged positively, they can accelerate towards the negative electrode under electric field and create a black composite on the Pt electrode. The OH⁻ generation together with other electrons and ions can form Ni and $Ni(OH)_2$ on Gr sheets. Based on all the above processes, we finally have Ni and $Ni(OH)_2$ deposited as crystalline layers on Gr flakes. All above processes are confirmed by the effect of initial $NiCl_2$/water concentration on exfoliation, as depicted in Figure 1c. In this figure, adding $NiCl_2$ results in higher production of Gr sheets, but below and above these values, final products do not show magnetic properties. Because OH⁻ and Cl⁻ are responsible for production of Gr flakes, therefore increasing molarity of $NiCl_2$ results in more exfoliated production content. As we are mainly interested in magnetic properties of our final products, we therefore investigate all of our samples made with 0.05, 0.075 and 0.1 M. In continue, we represent a comprehensive study of products and investigate their conductivity, magnetization and MI measurements and discuss the mechanism of MI theoretically.

## 2.2. Characterization

The crystalline structure of samples was characterized using X-ray diffractometer (XRD) with Cu Kα (λ = 0.154 nm) radiation. Fourier transform infrared (FT-IR) spectra were recorded via a Bruker (Tensor 27) FT-IR spectrometer with resolution of 1 cm$^{-1}$ in transmission mode at room temperature. X-ray photoelectron spectroscopy (XPS) was done in an ESCA/AES system equipped with a concentric hemispherical analyzer (CHA, Specs model EA10 plus). The size and morphology of elements were observed by tunneling electron microscopy (TEM-Philips model CM120). Room temperature magnetization measurements were done via vibrating sample magnetometer (Meghnatis Daghigh Kavir Co.). Current-voltage (I-V) measurements were done by two probe method using Keithley (model 2450) as sourcemeter.

## 2.3. Magnetoimpedance (MI) Measurement

Amorphous Co-based ribbons $Co_{68.15}Fe_{4.35}Si_{12.5}B_{15}$ (1 mm width, 40 mm length and ~20 μm thickness) is prepared by a conventional melt-spinning technique. The Gr/Ni composite with different molarities of 0.05, 0.075 and 0.1 M of $NiCl_2$ was drop coated on the two surface sides of the ribbon at room temperature. Evaporation of water and nanoparticle solution was required before the data acquisition process start. To measure the MI response of the samples, an external magnetic field produced by a solenoid was applied along the ribbon axis and the impedance was measured by means of the four-point probe method. The ac current passed through the longitudinal direction of the ribbon with different frequencies was supplied by a function generator (GPS-2125), with a 50 Ω resistor in the circuit. The impedance was evaluated by measuring the voltage and current across the sample using a digital oscilloscope (GPS-1102B). The MI ratio is defined as

$$MI\% = \frac{Z(H_e) - Z(H_{e_{max}})}{Z(H_{e_{max}})} \times 100; \tag{2}$$

where $Z$ refers to the impedance as a function of the external field $H_e$. The $H_{e_{max}}$ is the maximum field applied to the samples during the MI measurement.

## 3. RESULT AND DISCUSSION

XRD analysis was employed to confirm the crystalline structures of Gr/Ni composites as shown in Figure 2a. The dominant diffraction peak of the initial graphite foil at 2θ = 26.31° indicates an interlayer *d*-spacing of 3.38 Å, while in the final samples the (002) diffraction peak is appeared at 26.16° with an interlayer *d*-spacing of 3.40 Å. There are some diffraction peaks that belong to Ni(OH)$_2$. One broad diffraction peak at around 2θ=13.45° and another one at 2θ = 33.2° represent the (001) and (100) planes of the β-phase, respectively (JCPDS No: 001-1047). While (110) and (111) planes of the α-phase appear with broad diffraction peak at 2θ = 36-39.3° (JCPDS No: 06-0075) [36]. Other two diffraction peaks with 2θ = 59° and 62.1° represent (110) and (111) planes of β-phase Ni(OH)$_2$. The diffraction peaks for Ni were also observed at 2θ = 44.6° and 52° which correspond to (111) and (200) planes, respectively (JCPDS No: 001-1260), indicating the presence of face centered cubic (fcc) Ni with lattice constant of 3.51 Å. By increasing the concentration of NiCl$_2$, less relative intensity of Ni peaks can be seen. The average size of the deposited Ni and Ni(OH)$_2$ crystals were calculated based on Scherrer equation and found to be 43 and 32 nm, respectively. Values are close to those observed in TEM images that will be explained later.

To elucidate the interaction of Ni nanoparticles (Ni NPs) with the Gr sheets, FTIR spectra were recorded and analyzed. FTIR is a useful technique to confirm that the nanoparticles are anchored to Gr surface, as has been shown by several groups [37,38]. Figure 2b shows the FTIR spectra of the graphite foil and produced samples. It shows a number of oxygen functionalities and this may be because of the residual functional groups in the Gr [39]. The absorption peaks at ~1633 cm$^{-1}$, ~1550 cm$^{-1}$ and ~1375 cm$^{-1}$ should be assigned to stretching vibrations (υ) of C=O. The presence of O-H is confirmed by the strong and broad band at 3446 cm$^{-1}$. The intensity of this peak is stronger for the synthesized Gr-Ni structures than the initial graphite foil which shows the effect of aqoueos soloution of experiment on generation of hydroxyl elements. The absorption peak at 1053 cm$^{-1}$ should be attributed to stretching vibrations (υ) of C-O-C [40] that disappears after exfolation in Gr-Ni structures. The bands at 2925 cm$^{-1}$ and 2852 cm$^{-1}$ are assigned to the asymmetric and symmetric vibrations of C-H, respectively [41,42]. Comparing the FTIR spectra of graphite foil with the final sample, the spectrum of Gr-Ni structures clearly exhibite a considerable redshift (particularly to the major vibration of C=C and –OH bonds) in the FTIR peaks, which may be because of the Ni bonded to the Gr layers [43]. The absorption bands at 651, 563 and 426 cm$^{-1}$ are attributed to the NiO and n (Ni–OH) vibration [43].

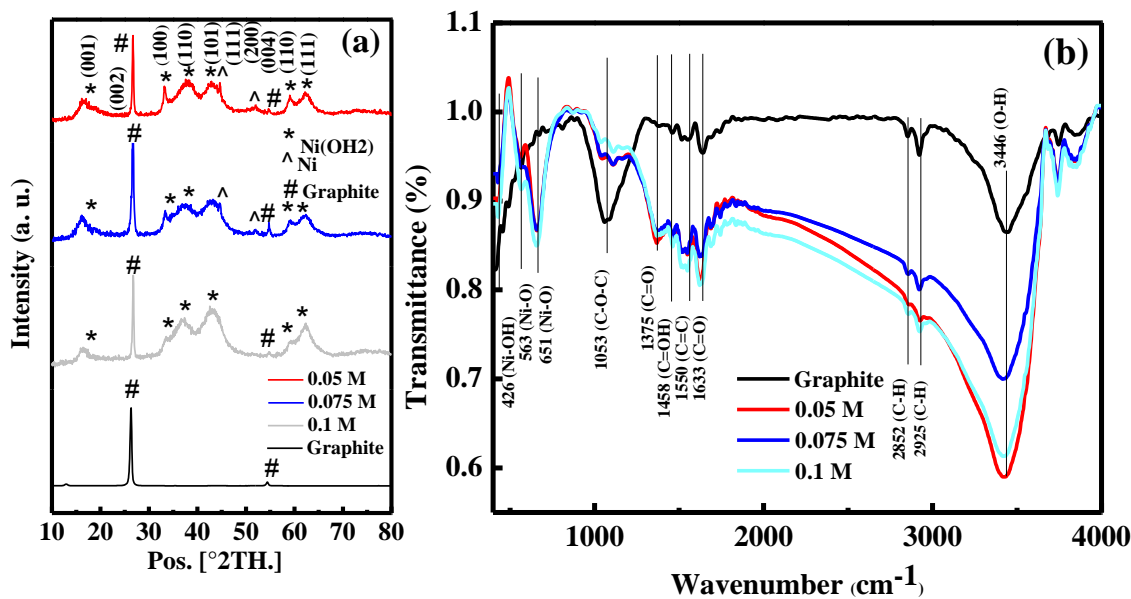

**Figure 2.** (a) X-Ray Diffraction (XRD) and (b) FTIR spectra of 0.1, 0.075 and 0.05 M samples and graphite foil.

High-resolution XPS spectra is used to probe the chemical compositions of the 0.05 and 0.1 M Gr/Ni samples (Figure 3a-d). The C 1s and Ni 2p peaks are decomposed into multi-peaks analyzing with distribution of C–C/C–O/C=O and Ni–O/ Ni–OH bonds. Peak positions and FWHM values are presented in Table 1. The peaks related to Ni 2p show presence of Ni(OH)$_2$ and Ni-oxides and no Ni. But Ni peaks were observed in the XRD measurements which reconfirms the presence of Ni compounds. This contrary is expected as the XPS is able to probe the surface of the materials only. Also, rather than the oxides at the surface of the Ni, there are some other covering compounds which cause the peak related to the Ni does not appear in the spectrum.

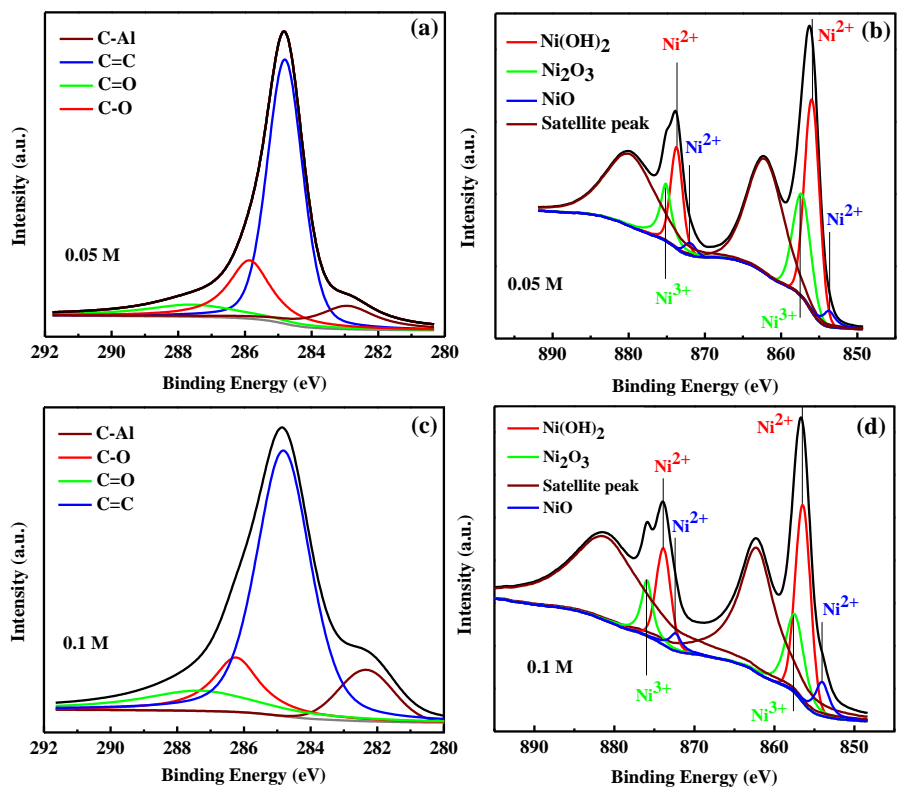

**Figure 3.** High-resolution XPS spectra of C 1s and Ni 2p core level of Gr/Ni sample of (a, b) 0.05 M and (c, d) 0.1 M.

**Table 1.** Peak positions and FWHM values from high-resolution XPS spectra of C 1s and Ni 2p core level of 0.05 and 0.1 M Gr/Ni sample.

| sample | C1s(C−Al) | | C1s(C=C) | | C1s(C−O/C−Cl) | | C1s(C=O) | |
|---|---|---|---|---|---|---|---|---|
| | B.E. (eV) | FWHM (eV) | B.E. (eV) | FWHM (eV) | B.E. (eV) | FWHM (eV) | B.E. (eV) | FWHM (eV) |
| 0.05 M | 282.5 | 2 | 284.8 | 1.3 | 285.9 | 1.7 | 287.6 | 4.1 |
| 0.1 M | 282.6 | 2.6 | 284.8 | 1.6 | 285.7 | 1.7 | 287.3 | 4.1 |
| sample | Ni2p$_{3/2}$(Ni$^{2+}$) | | Ni2p$_{3/2}$(Ni$^{3+}$) | | Ni2p$_{1/2}$(Ni$^{2+}$) | | Ni2p$_{1/2}$(Ni$^{3+}$) | |
| | B.E. (eV) | FWHM (eV) | B.E. (eV) | FWHM (eV) | B.E. (eV) | FWHM (eV) | B.E. (eV) | FWHM (eV) |
| 0.05 M | 855.9 - 853.7 | 2.3 - 1.9 | 857.4 | 2.7 | 873.6 - 871.9 | 1.8 - 1.9 | 875.1 | 1.7 |
| 0.1 M | 856.4 - 854 | 2.3 - 2 | 857.4 | 2.7 | 873.9 - 872.4 | 2.2 - 1.8 | 875.9 | 1.6 |

The area ratios allow calculating relative proportions of atoms in each binding for a given element. The concentration ratio of Ni:C was estimated using the following relation for 0.05 and 0.1 M samples:

$$\frac{C_{Ni}}{C_C} = \frac{I_{Ni}}{S_{Ni}} \times \frac{S_C}{I_C} \qquad (3)$$

where C, I and S are the elemental concentrations, XPS peak area and corresponding sensitivity factors. The relative surface ratio of $\frac{C_{Ni}}{C_C}$ for 0.1 M sample is 0.36 which increases to 0.51 for 0.05 M sample. These results indicate that Ni was enriched on the surface of carbon. Also $\frac{C_{Ni}}{C_O}$ for 0.1 M sample is 0.71 which increases to 0.85 for 0.05 M sample and $\frac{C_C}{C_O}$ for 0.05 M sample is 1.66 and increases to 1.93 for 0.1 M sample.

TEM images of samples are shown in Figure 4. Hexagonal structures appearing in Figure 4a-c are presumed to be related to Ni(OH)$_2$ nano-crystals, which their crystallization at β-hexagonal phase was identified from our XRD data. This Ni-based nano-crystals are separated by dashed lines in Figure 4c. They have also different alignments like rolled up structures, which are shown by an arrow in Figure 4c that occurs with more contents with increasing molarities, i.e. 0.1 M sample. Comparing all features dictates small thickness of such crystal, as they look to have thin planar geometry. In Figure 4b, d we see crumpled Gr and nano-crystals stick to it for 0.75 M sample. The nano-crystals (Ni, Ni(OH)$_2$, Ni-oxides) are randomly distributed on Gr sheets which are supposed to be the reason of their superparamagnetic-like response (presented at continue). TEM results confirm the size of Ni-based nano-crystals obtained by XRD data. The brightness of Gr sheets appearing in this image shows their small thickness. The image in Figure 4d shows micron size formation of Gr sheets which is in agreement with DLS results. DLS measurements showed the size of particles to be about 1.2 μm. SAED pattern in Figure 4e, measured for a randomly selected Gr sheet of the 0.05 M sample, represents the polycrystalline nature of the Gr.

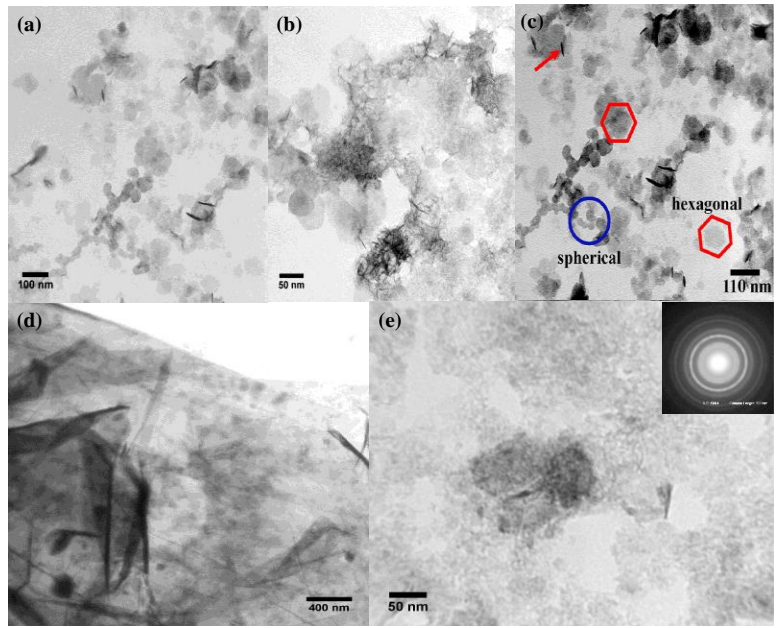

**Figure 4.** TEM images of (a) 0.1 M, (b) 0.075 M, (c) 0.05 M samples and (d) micron boundary size of Gr sheets and (e) SAED pattern of the 0.05 M sample.

To investigate the effect of the molarity of NiCl$_2$ on the magnetic properties of the Gr/Ni composite, the magnetic hysteresis loops of the 0.05, 0.075 and 0.1 M samples were recorded at room temperature. The VSM results shown in Figure 5a, represent superparamagnetic nature of the samples. By decreasing the molarity of NiCl$_2$ the magnetization of the product increases (see inset of Figure 5a ). The coercivity for all of the samples was very small, indicating that the majority of the MNPs are in a superparamagnetic state. XRD results showed the lower consentration of Ni in samples synthesized by higher NiCl$_2$ molarities. The probability of the formation of Ni or Ni(OH)$_2$ depends on the consentration of OH$^-$ at the cathode. It means that the more the molarity of the NiCl$_2$ results in more generation rate of OH$^-$ and therefore the more ratio of Ni(OH)$_2$ to Ni would be gained. Consecuently, by controlling the molarity of the solution, the consentration of ferromagnetic Ni and so the magnetization will be controlled. The XPS results are compared with magnetization which indicates that the concentration of Ni to Ni(OH)$_2$ is more in 0.05 M sample and caused a higher magnetization. This also confirms that the conductivity of 0.05 M sample is more than that of 0.1 M sample which has been seen by our I-V experiment. For I-V measurements, silver paste coated pellets of the samples were used. I-V plots are linear, showing conductive behavior of all samples (Figure 5b). Resistivity of samples was determined by linear fitting of I-V plots. Resistivity increases from 0.05 M sample to 0.1 M sample which is shown in the inset of Figure 5b. According to the aforementioned growth mechanism, we can interpret that as the molarity of NiCl$_2$ in solution increases, there is more OH$^-$ generation on the cathode and more hydroxide elements in Gr, and hence more resistivity in samples. XPS and XRD data show more hydroxide densities and lower Ni elements for samples prepared from solutions with higher molarities of NiCl$_2$.

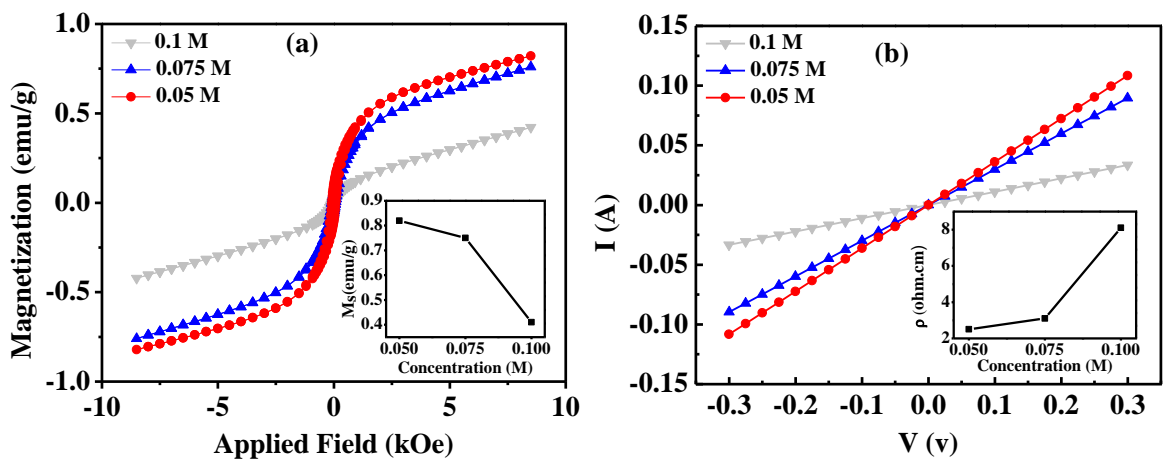

**Figure 5.** (a) VSM magnetization plots of 0.05 M, 0.075 M and 0.1 M samples. Inset shows the saturation magnetization of samples versus concentration. (b) Current-Voltage (I-V) curve measurement for 0.05 M, 0.075 M and 0.1 M samples and inset shows the resistivity ($\rho$) of samples versus concentration.

## 3.1. Magnetic Gr/Ni Detection by Magnetoimpedance Measurement: Testing of Biosensor

In the biosensor sector, the detection of functionalized magnetic nanoparticles has become a current active research issue [22,44]. Besides high sensitivities, the nanoparticle detector should also display low power consumption, small size, quick response, stability of operation parameters, resistance to aggressive medium, and low-cost. Different sensing techniques are employed for the nanoparticle detection: anisotropic magnetoresistance [45], spin-valves [46], giant magnetoresistance [47] and magnetoimpedance (MI) [48,49]. Among them, biosensors based on the so-called MI effect have been proposed as an alternative procedure for the detection of magnetic nanoparticles.

To evaluate the potential applications in sensing devices, we study the magnetic field and frequency dependences of MI response of a Co-based ferromagnetic ribbon and used this prototype as a background in order to determine the effects of the stray fields introduced by the magnetic Gr/Ni nanoparticles. By applying an ac charge current with frequency $f$ to the samples, we investigate how the impedance of the ribbon changes as a function of the external magnetic field. We perform field sweep impedance measurements at an arbitrary frequency of $f = 10$ MHz for ribbon (as-cast) and ribbon drop coated by Gr/Ni nanoparticle with three different molar ratios of 0.05, 0.075 and 0.1, and a current of $I = 66$ mA applied to the samples. Schematic illustration of the measurement setup for the MI response and the structural features of the MI based sensor is presented in Figure 6a. The lower the molarity of the samples, the higher the content of the Ni is in the sample. Each of the Ni nanoparticles can produce the stray field on the surface of the ribbon and thereby affect the magnetic anisotropy and permeability of the ribbon. As can be seen from Figure 6b, the MI ratio for the bare ribbon has the smallest value and increases for the drop-coated samples. The maximum values of the MI ratio are 201%, 218%, 238% and 275% for the bare ribbon and the ribbon with the drop-coated Gr/Ni composite with molar ratios of 0.1, 0.075 and 0.05, respectively. The inset of Figure 6b shows an enlarged portion of the low-field MI curves. One can see that the anisotropy field is 1.5 Oe for the bare ribbon while for the drop-coated sample, the amount of the anisotropy field decreases and reaches to 1 Oe. When the Ni nanoparticles are affected by the external field, they can produce a measurable stray field. As a result, presence of the Gr/Ni composite leads to a sizable increase in the MI ratio near the anisotropy field and the displacement of the MI curve. This finding is of practical importance, as the Gr/Ni composite can be better used as a magnetic biomarker for applications in medical diagnosis. Specially in the cases which Gr plays the role of bio marking [50]. The increase in the MI ratio due to the presence of the Ni nanoparticles or the Gr/Ni nanocomposite can be explained by considering the disturbance of the applied dc longitudinal and ac transverse fields caused by the presence of the stray fields of the Ni nanoparticles on the surface of the ribbon. In order to evaluate difference between the MI ratio of the bare ribbon and Gr/Ni composite covered ribbon, we measured the MI at different frequencies ranging from 1 MHz to 15 MHz. The maximum MI ratio of all samples versus frequency is plotted in Figure 6c. It is noted in Figure 6c that for all investigated samples, with increasing frequency, the maximum MI ratio first increases, reaches a maximum at a particular frequency (10 MHz), and then decreases for higher frequencies. This trend can be interpreted by considering the relative contributions of DW motion and moment rotation to the transverse magnetic permeability and hence to the MI [20,23,51]. Note that as frequency increases well above 100 kHz, the contribution of DW motion is damped due to the presence of the eddy current and moment rotation becomes dominant [20]. Thus, the MI ratio decreases at high frequencies.

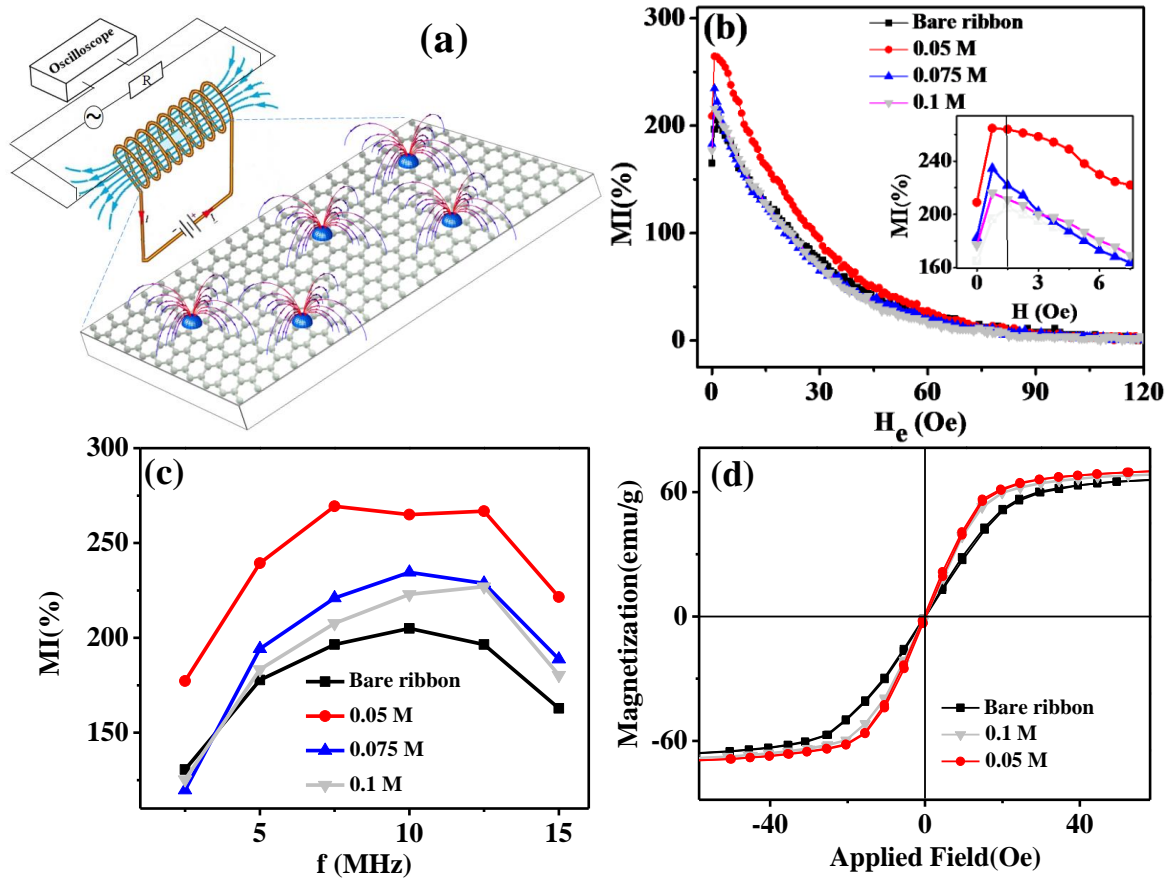

**Figure 6.** (a) Schematic illustration of the measurement setup for the MI response and the structural conditions in the MI sensor when ribbon is coated with Gr/Ni composites. The influence of the Gr/Ni layer on the MI is related to stray fields induced by magnetic Ni nanoparticles. The stray fields change the magnetization distribution in the magnetic ribbon and affect the permeability and the MI effect. (b) The MI ratio as a function of applied magnetic field for bare ribbon and ribbon drop coated by Gr/Ni nanoparticle with three different molar ratios. The inset shows an enlarged portion of the low-field MI curves. (c) Frequency dependence of the impedance response of all samples. (d) Magnetic hysteresis loops of bare ribbon and ribbon drop coated by Gr/Ni nanocomposite with molar ratios of 0.05 and 0.1.

Finally, to investigate the effect of the drop-coated Gr/Ni composite on the magnetic properties of the ribbons, conventional VSM method was used to obtain the magnetic hysteresis loops. Figure 6d shows the magnetic hysteresis loops of bare ribbon and ribbon drop coated by Gr/Ni composite with molar ratios of 0.05 and 0.1. It can be seen that the loops are all thin and narrow, and magnetization was saturated at a small applied field, indicating their soft ferromagnetic characteristics. According to the Figure 6d, after coating, the saturation field decreases and the differential magnetic permeability of the sample which is proportional to the magnetization slope has increased dictating a weak field sensitive and magnetically softer state. In order to understand the obtained experimental results, we proposed and developed the following electromagnetic model.

## 4. MODEL FOR IMPEDANCE OF RIBBON/GrNi STRUCTURES

In order to describe the experimental results, we propose a model for the MI response of the ribbon/ GrNi structures. MI response can be found by means of a solution of Maxwell equations for the electromagnetic fields and Landau–Lifshitz–Gilbert (LLG) equation for the magnetization dynamics with appropriate boundary conditions at the interfaces between the layers and outer surfaces. The influence of the GrNi layer on the MI is related to stray fields induced by magnetic Ni nanoparticles. The stray fields change the magnetization distribution in the magnetic ribbon and affect the permeability and MI effect. To describe qualitatively the influence of stray fields on the MI, we assume that the GrNi layer generates a spatially uniform effective field $H_s$ in the film. The value of $H_s$ is assumed to be proportional to the concentration of MNPs in the GrNi, since the GrNi saturation magnetization increases linearly (see Figure 5a) with the concentration of nanoparticles. It is assumed that the effective stray field $H_s$ has the opposite direction with respect to the magnetization vector in the GrNi layer. Also, we assume that the values of the permeability in the magnetic layer are determined by the magnetization rotation only because the domain-wall motion is strongly damped at sufficiently high frequencies (above 100 kHz) [20].

A schematic of the ribbon/ GrNi structures and the coordinate system used for the analysis is shown in Figure 7. There are three regions ''0,'' ''1,'' and ''2'' denoting the magnetic ribbon layer, the GrNi layer, and air. The film structure having length $l$ and width $w < l$ and consists of thickness $t_0$ and $t_1$ for the ribbon and Gr/Ni layer, respectively. The structure is subjected to an alternating driving field $E = E_0 e^{-i\omega t}$, and an external magnetic field $H_e$ is parallel to the long side of the sample (y-axis). It is assumed that the film length and width are much larger than its thickness. Introducing the scalar and vector potentials, $\varphi$ and $\boldsymbol{A}$, Maxwell's equations can be expressed as ($\boldsymbol{\nabla} \cdot \boldsymbol{A} = 0$ is used)

$$\nabla^2 A_j(r,t) = \mu_j \sigma_j \left[ \nabla \varphi_j(r,t) + \frac{d}{dt} A_j(r,t) \right]$$
$$\nabla^2 \varphi_j(r,t) = 0 \qquad (4)$$

where $j = 0, 1$ corresponds to regions "0" and "1" and $\sigma_j$, $\mu_j$ corresponds to conductivity and permeability of the magnetic ribbon for $j = 0$ and Gr/Ni layer for $j = 1$. The general solutions for the vector potential in the three regions can be expressed as [52]

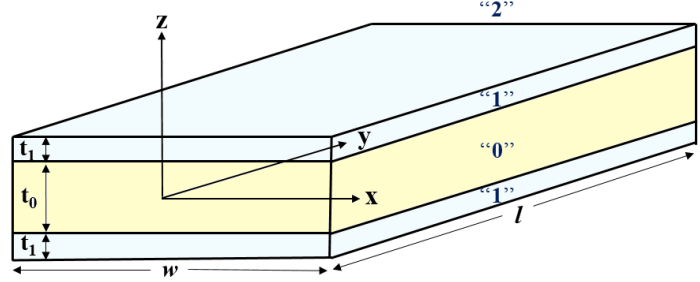

**Figure 7.** Schematic of the ribbon/ GrNi structures and the coordinate system.

$$A_0 = \frac{iE_0}{\omega} [B_0 \cosh(\alpha_0 z) + C_0 \sinh(\alpha_0 z) - 1] \qquad (5)$$

$$A_1 = \frac{iE_0}{\omega} [B_1 \cosh(\alpha_1 z) + C_1 \sinh(\alpha_1 z) - 1] \qquad (6)$$

$$A_2 = \frac{iE_0}{\omega} B_2 \left[ \frac{l}{w} \ln\left(\frac{r+w}{r-w}\right) - \frac{2z}{w} \arctan\left(\frac{w\ l}{2y\ r}\right) + \ln\left(\frac{l+r}{\sqrt{w^2 + 4z^2}}\right) \right] \qquad (7)$$

Where, $r = (w^2 + 4z^2 + l^2)^{\frac{1}{2}}$, $\alpha_k = (1+i)(\frac{\omega \mu_j \sigma_j}{2})^{\frac{1}{2}}$ and due to symmetry along the Z axis, $A_0(Z) = A_0(-Z)$, the constant of $C_0$ is zero. The boundary conditions allow one to find the constants $B_0$, $B_1$, $C_1$, and $B_2$ in Equations (5)–(7) and describe completely the distribution of the vector potential (see Appendix). When the potential distribution is obtained, since the magnitude of the driving electric field is constant, the impedance can be obtained as the proportionality factor between the voltage and the total current in the device:

$$Z = \frac{lE_0}{2w} \left( \int_0^{t_0/2} J_0(z)\, dz + \int_{t_0/2}^{t_0/2 + t_1} J_1(z) dz \right)^{-1} \qquad (8)$$

where,

$$J_0 = E_0 \sigma_0 B_0 \cosh(\alpha_0 z) \qquad (9)$$
$$J_1 = E_0 \sigma_1 [B_1 \cosh(\alpha_1 z) + C_1 \sinh(\alpha_1 z)] \qquad (10)$$

Finally, we obtain the impedance Z of the film with the Gr/Ni layer:

$$Z = \frac{l}{2w}\frac{1}{\gamma'}\frac{\cosh\left(\frac{\alpha_1 t_0}{2}\right)\cosh(\alpha_1 t)\left[\left(\tanh(\alpha_1 t)-\tanh\left(\frac{\alpha_1 t_0}{2}\right)\right)(\gamma+\beta\gamma'\tanh\left(\frac{\alpha_0 t_0}{2}\right))+(1-\tanh\left(\frac{\alpha_1 t_0}{2}\right)\tanh(\alpha_1 t))(\gamma'+\beta\gamma\tanh\left(\frac{\alpha_0 t_0}{2}\right))\right]}{\frac{2\sigma_0}{\alpha_0}\tanh\left(\alpha_0\frac{t_0}{2}\right)+\frac{2\sigma_1}{\alpha_1}[\sinh(\alpha_1 t_1)+\beta\tanh\left(\frac{\alpha_0 t_0}{2}\right)\cosh(\alpha_1 t_1)-\beta\tanh\left(\frac{\alpha_0 t_0}{2}\right)]} \quad (11)$$

Where, $t = \frac{t_0}{2} + t_1$, $\gamma = 1 + \ln\frac{2l}{w}$, $\gamma' = \frac{\pi}{w}\frac{\mu_1}{\mu_2\alpha_1}$ and $\beta = \left(\frac{\mu_1\sigma_0}{\mu_0\sigma_1}\right)^{\frac{1}{2}}$.

### 4.1. Effect of Gr/Ni Layer on Ribbon Permeability

The MI response of the ribbon is controlled by the transverse magnetic permeability. The transverse permeability depends on many factors, such as the domain structure, anisotropy axes distribution, mode of the magnetization variation, and so on. Since in the experiment the current frequencies are sufficiently high, the value of the transverse permeability in the ferromagnetic layers is governed by the magnetization rotation. This approximation is valid at sufficiently high frequencies (>100 KHz), when the domain-wall motion is damped [20]. We also suppose that the ferromagnetic layers have in-plane uniaxial anisotropy and the direction of the anisotropy axis is close to the transverse one.

The magnetization distribution in the ferromagnetic layers can be found by minimizing the free energy. Taking into account the effective stray field, $H_s$, the minimization procedure results in the following equation for the equilibrium magnetization angle, $\theta$:

$$H_a \sin(\theta - \psi)\cos(\theta - \psi) - H_s \sin(\theta - \varphi) - H_e \cos\theta = 0 \quad (12)$$

Here, $H_a$ is the anisotropy field in the ferromagnetic layers and $\psi$ is the deviation angle of the anisotropy axis from the transverse direction.

In the magnetic susceptibility model, magnetization dynamics is governed by the Landau-Lifshitz-Gilbert equation, given by

$$\frac{\partial \mathbf{m}}{\partial t} = -\gamma \mathbf{m} \times \mathbf{H}_{eff} + \alpha \mathbf{m} \times \frac{\partial \mathbf{m}}{\partial t} \quad (13)$$

where $\mathbf{m}$ is vector magnetization, $\mathbf{H}_{eff}$ is the effective field, $\gamma$ is the electron gyromagnetic ratio and $\alpha$ is the Gilbert damping constant. The solution of the linearized LLG equation results in the following expression for the transverse permeability in the ribbon layer [53]:

$$\mu_0 = 1 + \frac{\omega_m[\omega_m + \omega_1 - i\alpha\omega]\sin^2\theta}{[\omega_m + \omega_1 - i\alpha\omega][\omega_2 - i\alpha\omega] - \omega^2} \quad (14)$$

Here $\omega_m = 4\pi\gamma M_s$, $M_s$ is the saturation magnetization of the ribbon, $\omega = 2\pi f$ is the angular frequency, and

$$\omega_1 = \gamma[H_a \cos^2(\theta - \psi) - H_s \cos(\theta - \varphi) + H_e \sin\theta] \quad (15)$$
$$\omega_2 = \gamma[H_a \cos2(\theta - \psi) - H_s \cos(\theta - \varphi) + H_e \sin\theta] \quad (16)$$

Thus, the MI response in the ribbon with a Gr/Ni layer can be calculated as follows. The first step is the determination of the transverse permeability in the ferromagnetic layer by using Eq. (14). The second step is substitution of Eq. 14 into Eq. 11 and then the calculation of the impedance Z of the film with the Gr/Ni layer. When the impedance Z is found, the MI response of the ribbon with Gr/Ni layer can be obtained by means of Equation (2).

## 4.2. MI Response of Ribbon/ GrNi Structures: Comparison of Experimental and Modelling Results

Figure 8a shows the field dependence of the MI ratio ΔZ/Z calculated at 10 MHz for the ribbon without Gr/Ni layer and the ribbon with Gr/Ni layer for different values of the effective stray field, $H_s$. Note that the results are presented only for the range of positive fields, since the MI ratio is symmetric with respect to the sign of the external field. The effective stray field increases with the concentration of MNPs in the Gr/Ni layer due to the growth of the Gr/Ni layer saturation magnetization. As a result, the MI ratio shifts toward low external fields with increasing $H_s$ and exhibits a dependence similar to the one observed in the experiment.

To analyze the variation of the MI effect, let us introduce the impedance field sensitivity, which is defined as follows

$$S = \frac{|\Delta Z|}{|\Delta H_e|} = \frac{Z(H_e = -H_a) - Z(H_e = 0)}{H_a} \quad (17)$$

Figure 8b presents the frequency dependence of the impedance sensitivity calculated by means of Eq. (17) for different values of the effective stray field, $H_s$. The field sensitivity increases at relatively low frequencies and attains a peak at f= 20 MHz. With a growth of the effective stray field, $H_s$, the position of the highest sensitivity shifts to lower frequencies.

The proposed model allows one to describe qualitatively the main features of the experimental results of the MI response of a ribbon with a Gr/Ni layer. However, some experimental results cannot be explained in the framework of the model. It is demonstrated that the contribution of the stray fields induced by MNPs in the Gr/Ni layer leads to a shift toward low external fields with an increase of $H_s$. It should be noted that the model does not describe the essential increase of the MI response with an increase in the concentration of MNPs in the Gr/Ni layer. The disagreement between theoretical and experimental results may be attributed to the fact that in the proposed model, the contribution of surface modification, i.e. roughness, is not considered. There are many experimental works that demonstrated the fact that reduction of roughness can lead to increase of the MI response [27,28,31,33]. The increase of MI response in the magnetic layer deposited on ribbon has been already explained according to modifications of the ribbon surface and the closure of magnetic flux paths of deposited ribbons with magnetic materials. Note also, that the simplified presentation of the stray fields created by the MNPs by means of the effective field, $H_s$ qualitatively describes the effect of the Gr/Ni layer on the MI of the ribbon. Therefore, to estimate the exact value of $H_s$, an approximate distribution of the stray fields should be found by means of a numerical solution for the magnetostatic equation.

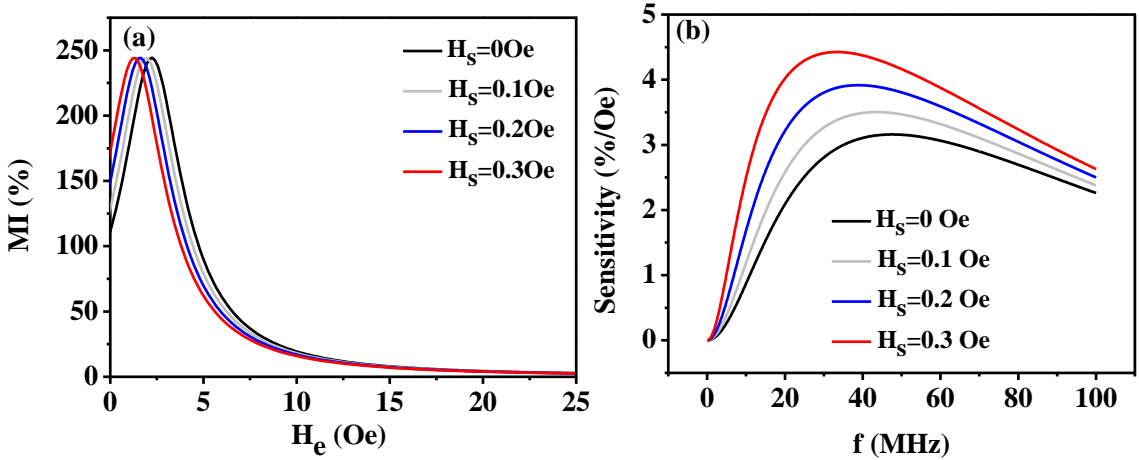

**Figure 8.** (a) The field dependence of the MI ratio ΔZ/Z calculated at f = $\frac{\omega}{2\pi}$ =10 MHz for the ribbon without GrNi layer and the ribbon with GrNi layer for different values of the effective stray field $H_s$. (b) The frequency dependence of the impedance sensitivity calculated by means of Eq. (17) for different values of $H_s$. Parameters used for calculations are $w = 8mm$, $l = 4cm$, $t_0 = 20\mu m$, $t_1 = 10\mu m$, $\sigma_0 = 5.81 \times 10^7 \frac{1}{\Omega m}$, $\sigma_1 = 10^3 \frac{1}{\Omega m}$, $M_s = 800$ Oe, $\theta = 0.75\pi$, $\varphi = -0.1\pi$, $\psi = 0.48\pi$, $\alpha = 0.1$, $H_a = 4$ Oe.

## 5. CONCLUSION

In summary, the present study demonstrates that electrochemically exfoliated Gr is a promising method for fabricating magnetic Gr sheets/Ni-nano-crystal composites in high quality. By changing the molarity of a NiCl$_2$ aqueous solution, different magnetic strengths of the final composite material were gained. Different molarities of 0.05, 0.075 and 0.1 M of NiCl$_2$ for fabrication of the samples results in different magnetization of the material. The higher molarity results in lower magnetization. In addition to the focus on the novel fabrication of materials, we further investigate the application of the synthesized materials in MI sensors with observed enhanced ratio and sensitivity when applied atop an MI sensor. The experimental observation of the MI response of our sensor shows higher MI% for the sensor coated with the Gr/Ni sample with a higher magnetization. A corresponding MI model was developed on the basis of the solution of linearized Maxwell equations and Landau-Lifshitz equation for the magnetization dynamics in order to understand the origin of the behavior of the MI response in the presence of Gr/Ni composite. The results can be considered for future developments of MI based sensors and biosensors.

**Appendix**

Boundary conditions require $A$ and $\frac{1}{\mu}\frac{dA}{dz}$ be continuous:

$$A_0 = A_1 \qquad z = \frac{t_0}{2}$$

$$\mu_1 \frac{d}{dz} A_0 = \mu_0 \frac{d}{dz} A_1$$

$$A_1 = A_2 \qquad z = \frac{t_0}{2} + t_1$$

$$\mu_2 \frac{d}{dz} A_1 = \mu_1 \frac{d}{dz} A_2 \tag{A1}$$

By substitution of Eq. (5-7) into Eq. (A1) with approximation of $A_2$ for $w \ll l$:

$$\begin{aligned}
B_0 \cosh(x_0) - B_1 \cosh(y_0) - C_1 \sinh(y_0) &= 0 \\
B_0 \beta \sinh(x_0) - B_1 \sinh(y_0) - C_1 \cosh(y_0) &= 0 \\
B_1 \cosh(z_0) + C_1 \sinh(z_0) - B_2 \gamma &= 1 \\
B_1 \sinh(z_0) + C_1 \cosh(z_0) + B_2 \gamma' &= 0
\end{aligned} \tag{A2}$$

where

$$x_0 = \frac{\alpha_0 t_0}{2} \qquad y_0 = \frac{\alpha_1 t_0}{2} \qquad z_0 = \alpha_1 t$$

$$t = \frac{t_0}{2} + t_1 \qquad \gamma = 1 + \ln\frac{2l}{w} \qquad \gamma' = \frac{\pi}{w}\frac{\mu_1}{\mu_2 \alpha_1}$$

Solving for the coefficients we obtain:

$$B_0 = \frac{1}{E}$$
$$B_1 = \frac{1}{E}(\cosh(x_0)\cosh(y_0) - \beta \sinh(x_0)\sinh(y_0))$$
$$C_1 = \frac{1}{E}(\beta \sinh(x_0)\cosh(y_0) - \sinh(y_0)\cosh(x_0))$$
$$E = \frac{1}{\gamma'}\cosh(x_0)\cosh(y_0)\cosh(z_0)\left[(\tanh(z_0) - \tanh(y_0))(\gamma + \beta\gamma' \tanh(x_0)) + (1 - \tanh(y_0)\tanh(z_0))(\gamma' + \beta\gamma \tanh(x_0))\right] \tag{A3}$$

## AUTHOR INFORMATION

**Corresponding Authors**

*M.R.H. e-mail: mrh.hajiali67@gmail.com
*M.M. e-mail: m-mohseni@sbu.ac.ir

**ORCID**

Mohammadreza Hajiali: https://orcid.org/0000-0001-5068-4437

Majid Mohseni: https://orcid.org/0000-0001-5626-533X

**Author Contributions**
The manuscript was written through contributions of all authors. All authors have given approval to the final version of the manuscript.

## ACKNOWLEDGMENTS

Support from Iran National Science Foundation (INSF) is acknowledged.